\documentclass{article}
\usepackage[table]{xcolor}
\usepackage{spconf,amsmath,amssymb,graphicx}
\usepackage{booktabs,array,tabularx,capt-of,balance}
\usepackage{tikz,pgfplots}
\usetikzlibrary{positioning,fit,calc}
\usepackage{microtype}
\usepackage[hidelinks]{hyperref}
\pgfplotsset{compat=1.18}

\hypersetup{pdftitle={SkillFormer: Skill-Decomposed Adaptation for Audio Language Models}}
\newcommand{\best}[1]{\textbf{#1}}
\newcommand{\second}[1]{\underline{#1}}
\newcolumntype{N}{>{\centering\arraybackslash}p{3.3em}}
\newcolumntype{A}{>{\centering\arraybackslash}p{4.2em}}
\newcommand{\paperfont}{\fontsize{9}{10.8}\selectfont}
\definecolor{tblNavy}{HTML}{243B53}
\definecolor{tblBand}{HTML}{EEF2F5}
\definecolor{tblStripe}{HTML}{F6F8FA}
\definecolor{tblOurB}{HTML}{EBF6F2}
\definecolor{tblOurInk}{HTML}{174E45}
\definecolor{tblRule}{HTML}{52687A}
\definecolor{tblGain}{HTML}{157951}
\definecolor{tblOurA}{HTML}{E9F4F1}
\newcommand{\hd}[1]{\textcolor{tblNavy}{\textbf{#1}}}
\newcommand{\grp}[1]{\textcolor{tblNavy}{\textit{#1}}}
\newcommand{\ours}[1]{\textcolor{tblOurInk}{\textbf{#1}}}
\newcommand{\gain}[1]{{\fontsize{7.2}{7.2}\selectfont\color{tblGain}$+$#1}}
\title{SkillFormer: Skill-Decomposed Adaptation\\for Audio Language Models}
\name{\begin{tabular}[t]{@{}c@{}}
Lee Seung-woo\textsuperscript{1}, Bowen Qi\textsuperscript{2}, Kim Min-jun\textsuperscript{3}, Jang Won-young\textsuperscript{3},
\end{tabular}}
\address{\textsuperscript{1}Pusan National University\quad%
\textsuperscript{2}Shanghai Jiao Tong University\quad%
\textsuperscript{3}Hanyang University}

\begin{document}
\ninept
\maketitle

\begin{abstract}
Audio language models must handle dozens of distinct skills, from pitch
comparison and speaker counting to musical tempo estimation and emotion
recognition. Joint training on all skills at once causes
interference: gains on one skill often come at the cost of another. We
propose \textbf{SkillFormer}, which decomposes audio understanding into
skill-specific low-rank adapters and composes them at inference time through
a learned router. The router examines the question to decide which adapters
to activate and how much weight each should carry, so that a pitch query
engages different parameters than a genre classification query. An
alternating training schedule updates each adapter on its own skill cluster
before jointly calibrating the router, preventing the gradient conflicts
that arise in standard multi-task optimization. SkillFormer adds fewer than
4\% of the base model's parameters and requires no changes to the audio
encoder or language backbone. Evaluated on three architecturally distinct
models across MMSU, MMAU-Pro, and MMAR, it raises the average accuracy by
2.5 to 4.1 points, with balanced gains across perception, reasoning, and
semantic subcategories.
\end{abstract}


\section{Introduction}
\label{sec:intro}
\vspace{-5pt}

Large audio language models
(LALMs)~\cite{qwen25,mimo,kimi,salmonn,ltuas,qwenaudio,speechgpt} answer
questions about speech, sound, and music through a single interface.
Benchmarks that probe individual competencies, however, expose a recurring
pattern: improving one cluster of skills during training frequently degrades
another~\cite{mmsu,mmaupro,mmar}. A model fine-tuned on sound event counting
may lose ground on speaker emotion, and vice versa. Broader evaluations
confirm that this skill interference is pervasive, spanning open-ended,
spatial, and compositional
queries~\cite{audiobench,airbench,dynamicsuperb,mmau}. The phenomenon
matters for downstream applications that require reliable performance on
specific competencies---paralinguistic analysis~\cite{wang2026parabridge},
nonverbal vocalization understanding~\cite{ni2026nv}, safety
assessment~\cite{wang2026voxsafebench}, and privacy
evaluation~\cite{wang2026voxprivacy} each depend on a narrow subset of audio
skills, and a model that excels on average may still fail on the subset that
matters.

Multi-task learning offers a natural framework: share a backbone while
specializing where
needed~\cite{multitask,pcgrad}. Mixture-of-experts
(MoE) architectures~\cite{moe,switchtransformer,mixtral,deepseekmoe} scale
this idea by routing each token through a sparse set of expert modules,
and parameter-efficient methods such as
LoRA~\cite{lora}, adapters~\cite{adapter}, AdaLoRA~\cite{adalora}, and
prefix tuning~\cite{prefix} make specialization practical even at the 7B
scale. In audio, however, the community has focused on other axes
of improvement. EvoAudio~\cite{wang2026evoaudio} and related
methods~\cite{af2,sari,audiozero,audiorubrics,aqattrl,sisda}
improve \emph{what} the model trains on through curriculum design, reward
shaping, and self-improvement. Recurrent latent reasoning
approaches~\cite{wang2026recurtrace,wang2026aural,cai2026thinking,feng2026coreloop}
change \emph{how deeply} the model reasons. Neither addresses \emph{which}
parameters the model activates for a given question.

We introduce \textbf{SkillFormer}, which fills this gap. The key idea is
simple: different audio skills should use different parameters. A
lightweight question-conditioned router selects a weighted combination of
$K$ low-rank adapters, each specialized to a cluster of related skills.
An alternating training schedule first updates each adapter on its own
cluster, then calibrates the router on mixed data, preventing the gradient
conflicts that degrade joint training~\cite{pcgrad}. The principle that
task-specific routing improves over monolithic processing has been validated
across modalities, from structured
reasoning~\cite{wang2026exploring,wang2025towards} to sequential
prediction~\cite{yong2025predicting,mtkd}, and
SkillFormer shows that it extends naturally to audio understanding.
MimicLM~\cite{feng2026mimiclm} and
Spatial-Omni~\cite{zhu2026spatial} illustrate that audio tasks are
diverse enough---spanning voice modelling, spatial perception,
and acoustic analysis---to benefit from per-skill specialization.
Strong audio representations from self-supervised
encoders~\cite{hubert,wavlm,wav2vec2}, supervised
models~\cite{whisper,ast,panns,conformer}, and contrastive
pretraining~\cite{clap,beats} provide the raw material; the question is
how to organize learning so that each skill gets the parameters it needs.
Reinforcement-based training~\cite{ppo,dpo,rlhf,grpo,dapo} and
chain-of-thought prompting~\cite{cot} have shown that targeted computation
improves language modelling; SkillFormer applies the same principle at the
parameter level for audio.

SkillFormer adds fewer than 4\% of the base model's parameters and
plugs into any LALM without modifying its encoder or language backbone.
We evaluate on Qwen2.5-Omni-7B, Kimi-Audio-7B-Instruct, and
MiMo-Audio-7B-Instruct across MMSU, MMAU-Pro, and MMAR. SkillFormer
raises the average accuracy by 2.5 to 4.1 points, with gains that
spread evenly across perception, reasoning, and semantic subcategories
rather than concentrating on one axis.

\begin{figure*}[t]
\centering
\begin{tikzpicture}[
  >=stealth, node distance=4mm and 5mm,
  box/.style={draw,rounded corners=2pt,minimum height=7.5mm,
    align=center,font=\paperfont\small,line width=0.35pt},
  adpt/.style={draw,rounded corners=1.5pt,minimum height=7mm,
    minimum width=13mm,align=center,font=\paperfont\scriptsize,
    line width=0.3pt},
  arr/.style={->,line width=0.55pt},
]
% --- Input ---
\node[box,fill=blue!6,minimum width=14mm] (wav) at (0,0) {Waveform};
\node[box,fill=blue!12,minimum width=18mm,right=5mm of wav]
  (enc) {Audio\\[-1pt]encoder};
% --- Features ---
\coordinate (feat) at ([xshift=12mm]enc.east);
% --- Adapter bank (stacked) ---
\node[adpt,fill=red!6] (a1) at ([xshift=32mm,yshift=14mm]enc.east)
  {Adapter 1\\[-1pt]{\tiny speech prosody}};
\node[adpt,fill=yellow!8] (a2) at ([xshift=32mm,yshift=4.5mm]enc.east)
  {Adapter 2\\[-1pt]{\tiny speakers \& dialogue}};
\node[adpt,fill=green!6] (a3) at ([xshift=32mm,yshift=-5mm]enc.east)
  {Adapter 3\\[-1pt]{\tiny sound events}};
\node[adpt,fill=purple!6] (ak) at ([xshift=32mm,yshift=-14.5mm]enc.east)
  {Adapter $K$\\[-1pt]{\tiny music}};
% dots between a3 and ak
\node[font=\paperfont\scriptsize,text=black!50]
  at ($(a3)!0.5!(ak)$) {$\vdots$};
% dashed border
\node[draw,densely dashed,rounded corners=3pt,
  inner xsep=4pt,inner ysep=3pt,
  fit=(a1)(a2)(a3)(ak),line width=0.45pt] (bank) {};
\node[font=\paperfont\footnotesize,anchor=south,yshift=1pt]
  at (bank.north) {Skill-Specific LoRA Bank};
% --- Router ---
\node[box,fill=orange!10,minimum width=20mm,
  below=10mm of bank] (router) {Router};
% --- Question ---
\node[box,fill=red!6,minimum width=20mm,
  left=8mm of router] (q) {Question $\mathbf{q}$};
% --- Gating output ---
\node[box,fill=orange!6,minimum width=20mm,minimum height=6mm,
  right=8mm of router,font=\paperfont\scriptsize]
  (gate) {$\alpha_1,\ldots,\alpha_K$};
% --- Aggregation ---
\node[box,fill=tblOurA,minimum width=16mm,
  right=8mm of bank] (sum) {$\sum \alpha_k$\\[-1pt]{\scriptsize merge}};
% --- LM ---
\node[box,fill=green!8,minimum width=18mm,right=8mm of sum]
  (lm) {Language\\[-1pt]model};
\node[box,fill=green!14,minimum width=14mm,right=5mm of lm]
  (ans) {Answer};
% --- Arrows: input flow ---
\draw[arr] (wav)--(enc);
\draw[arr] (enc)--
  node[above,font=\paperfont\scriptsize]{$\mathbf{H}$} (feat);
% fan out to adapters
\draw[arr] (feat)|-(a1.west);
\draw[arr] (feat)|-(a2.west);
\draw[arr] (feat)|-(a3.west);
\draw[arr] (feat)|-(ak.west);
% adapters to sum
\draw[arr] (a1.east)--(a1.east-|sum.west);
\draw[arr] (a2.east)--(a2.east-|sum.west);
\draw[arr] (a3.east)--(a3.east-|sum.west);
\draw[arr] (ak.east)--(ak.east-|sum.west);
% router flow
\draw[arr] (q)--(router);
\draw[arr] (router)--(gate);
\draw[arr] (gate.north)--(gate.north|-sum.south);
% output
\draw[arr] (sum)--(lm);
\draw[arr] (lm)--(ans);
\end{tikzpicture}
\vspace{-3pt}
\caption{SkillFormer routes audio features through a bank of $K$
skill-specific LoRA adapters. A question-conditioned router produces
gating weights $\alpha_1,\ldots,\alpha_K$; the adapter outputs are merged
and passed to the language model.}
\label{fig:overview}
\vspace{-5pt}
\end{figure*}
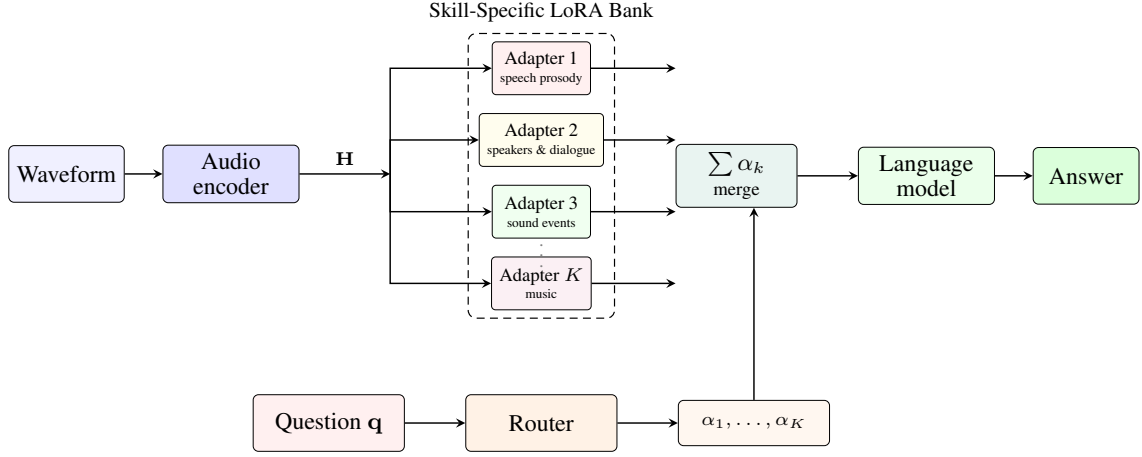

\vspace{-5pt}
\section{Skill-Decomposed Adaptation}
\label{sec:method}
\vspace{-5pt}

SkillFormer inserts a bank of $K$ low-rank adapters between the audio
encoder and the language model (Fig.~\ref{fig:overview}). A router
examines the question to decide which adapters to activate, and an
alternating schedule trains each adapter on its own skill cluster before
jointly calibrating the router.

\vspace{-3pt}
\subsection{Skill-Specific LoRA Adapters}
\vspace{-3pt}

Given audio encoder features $\mathbf{H}\in\mathbb{R}^{L\times d}$, each
adapter $k\in\{1,\ldots,K\}$ applies a low-rank
transformation~\cite{lora}:
\begin{equation}
\Delta\mathbf{H}_k = \mathbf{H}\,\mathbf{B}_k\,\mathbf{A}_k,
\qquad \mathbf{B}_k\in\mathbb{R}^{d\times r},\;
       \mathbf{A}_k\in\mathbb{R}^{r\times d},
\label{eq:lora}
\end{equation}
where $r\ll d$ is the adapter rank. Each adapter shares $r$ and the
projection dimensions but learns its own weights, so adapter~1 may
specialize in pitch and rate while adapter~3 handles sound events.
$\mathbf{B}_k$ is initialized from $\mathcal{N}(0,\sigma^2)$ and
$\mathbf{A}_k$ is initialized to zero, ensuring that the adapter
output is zero before training and the base model's behaviour is
preserved~\cite{lora}. A dropout of 0.05 is applied between $\mathbf{B}_k$
and $\mathbf{A}_k$ during training. The
adapters sit in parallel; their outputs are merged before entering the
language model.

\vspace{-3pt}
\subsection{Question-Conditioned Router}
\vspace{-3pt}

The router produces a sparse gating vector from the question.
It encodes the question tokens $\mathbf{q}\in\mathbb{R}^{T\times d}$
into a single vector $\bar{\mathbf{q}}$ via mean pooling, then applies a
two-layer MLP followed by a top-$n$ softmax:
\begin{equation}
\boldsymbol{\alpha} = \operatorname{Top\text{-}n\text{-}Softmax}
  \!\bigl(\mathbf{W}_2\,\operatorname{GELU}(\mathbf{W}_1\,\bar{\mathbf{q}}
  + \mathbf{b}_1) + \mathbf{b}_2\bigr),
\label{eq:router}
\end{equation}
where $\boldsymbol{\alpha}\in\mathbb{R}^K$ has at most $n$ nonzero entries
($n{=}2$ in all experiments). Sparsity keeps inference cost independent of
$K$~\cite{switchtransformer}. The merged adapter output is
\begin{equation}
\Delta\mathbf{H} = \sum_{k=1}^{K}\alpha_k\,\Delta\mathbf{H}_k,
\qquad
\hat{\mathbf{H}} = \mathbf{H} + \Delta\mathbf{H},
\label{eq:merge}
\end{equation}
and $\hat{\mathbf{H}}$ is passed to the language model backbone.

Including the question in the routing decision is critical: the same audio
clip paired with ``What language is being spoken?'' and ``Is the second
speaker's pitch higher than the first?'' should activate different adapters.
Audio-only routing would force both queries through the same path.

\vspace{-3pt}
\subsection{Skill Clustering}
\vspace{-3pt}

The $K$ skill clusters are derived from the training data before any
adapter is trained. Each training question is embedded with the language
model's own tokenizer, and $k$-means groups the embeddings into $K$
clusters. In our experiments $K{=}6$ produces clusters that align well
with intuitive skill families: speech prosody, speakers and dialogue,
speech content, sound events, music, and multi-clip scenes. This matches
the question taxonomy used by the EvoAudio tool
library~\cite{wang2026evoaudio}. Each cluster defines which training items
update which adapter.

\vspace{-3pt}
\subsection{Alternating Training}
\vspace{-3pt}

Simultaneous gradient updates from different skills can
conflict~\cite{pcgrad,multitask}, erasing gains on one skill while
optimizing another. SkillFormer avoids this with a two-phase alternating
schedule.

\textbf{Phase A --- adapter specialization.} For $T_A{=}500$ steps per
adapter, each adapter $k$ is updated on its own cluster with all other
adapters and the router frozen. The base audio encoder and language model
stay frozen throughout. This phase uses a learning rate of
$5{\times}10^{-5}$ with the Adam optimizer~\cite{adam} and cosine decay.

\textbf{Phase B --- router calibration.} For $T_B{=}2\,000$ steps, the
router and all adapters are trained jointly on a balanced sample across
all clusters, with the same learning rate reduced to $1{\times}10^{-5}$.
This phase teaches the router to compose adapters and allows cross-skill
adjustments. The language model parameters are unfrozen in this phase.

The full schedule runs phase~A once and phase~B once. Training data
consists of the same 50\,000 verifiable QA pairs from the EvoAudio tool
library~\cite{wang2026evoaudio}, drawn from
LibriSpeech~\cite{librispeech}, FSD50K~\cite{fsd50k},
AudioSet~\cite{audioset}, MELD~\cite{meld}, and synthesis via
Qwen3-TTS~\cite{tts}, LeVo~\cite{levo}, and
Scaper~\cite{scaper}. Each adapter $k$ is trained on roughly $50{,}000/K$
items in phase~A, while phase~B uses all 50\,000.

\begin{table*}[t]
  \centering
  \caption{Accuracy (\%) on three audio-understanding benchmarks.
  Avg.\ is the mean of the three All scores.
  SFT fine-tunes a single LoRA adapter on all data.
  Multi-LoRA keeps all $K{=}6$ adapters active with uniform weights.
  For MMAU-Pro, the three category columns do not cover every scored item,
  so All is not their average. \textbf{Bold} and \second{underline} mark
  the best and second best per backbone.}
  \label{tab:main}
  \centering
  \fontsize{9}{9.4}\selectfont
  \setlength{\tabcolsep}{1.5pt}
  \renewcommand{\arraystretch}{0.9}
  \arrayrulecolor{tblRule}
  \begin{tabular*}{\textwidth}{@{\extracolsep{\fill}}l*{12}{N}A@{}}
  \toprule
  & \multicolumn{3}{c}{\hd{MMSU $\uparrow$}} & \multicolumn{4}{c}{\hd{MMAU-Pro $\uparrow$}}
  & \multicolumn{5}{c}{\hd{MMAR $\uparrow$}} & \\
  \cmidrule(lr){2-4}\cmidrule(lr){5-8}\cmidrule(lr){9-13}
  \hd{Method} & \hd{Percep.} & \hd{Reason.} & \hd{All} & \hd{Speech} & \hd{Sound} & \hd{Music} & \hd{All} & \hd{Signal} & \hd{Percep.} & \hd{Seman.} & \hd{Culture} & \hd{All} & \hd{Avg. $\uparrow$} \\
  \midrule
  \rowcolor{tblBand}
  \multicolumn{14}{@{}l}{\grp{Qwen2.5-Omni-7B}} \\
  \addlinespace[1.3pt]
  Base & 44.5 & 79.4 & 61.4 & 57.8 & 45.3 & 64.0 & 56.4 & 55.8 & 54.0 & 67.2 & 58.9 & 60.2 & 59.3\phantom{\gain{0.0}} \\
  SFT & 47.0 & 80.3 & 63.2 & 59.1 & 47.0 & 65.4 & 57.5 & 56.5 & 55.3 & 67.8 & 59.7 & 61.0 & 60.6\gain{1.3} \\
  Multi-LoRA & \second{49.5} & \second{80.7} & \second{64.7} & \second{59.9} & \second{48.4} & \second{66.3} & \second{58.5} & \second{57.3} & \second{55.6} & \second{68.0} & \second{60.5} & \second{61.8} & \second{61.7}\gain{2.4} \\
  \rowcolor{tblOurB}
  \ours{SkillFormer} & \best{52.5} & \best{81.3} & \best{66.5} & \best{61.5} & \best{49.8} & \best{67.3} & \best{60.4} & \best{58.1} & \best{57.2} & \best{68.9} & \best{62.0} & \best{63.3} & \best{63.4}\gain{4.1} \\
  \midrule
  \rowcolor{tblBand}
  \multicolumn{14}{@{}l}{\grp{Kimi-Audio-7B-Instruct}} \\
  \addlinespace[1.3pt]
  Base & 39.0 & 74.3 & 56.0 & 58.9 & 37.1 & 56.1 & 49.6 & 46.5 & 47.3 & 63.1 & 51.1 & 54.3 & 53.3\phantom{\gain{0.0}} \\
  SFT & 41.5 & 75.4 & 58.0 & 60.2 & 39.4 & 57.5 & 51.1 & 48.0 & 48.7 & 64.0 & 52.6 & 55.4 & 54.8\gain{1.5} \\
  Multi-LoRA & \second{43.0} & \second{75.9} & \second{59.0} & \second{61.0} & \second{40.3} & \second{58.2} & \second{51.9} & \second{48.8} & \second{49.2} & \second{64.5} & \second{53.0} & \second{56.1} & \second{55.7}\gain{2.4} \\
  \rowcolor{tblOurB}
  \ours{SkillFormer} & \best{44.6} & \best{76.8} & \best{60.3} & \best{62.1} & \best{42.8} & \best{59.5} & \best{53.6} & \best{50.9} & \best{50.5} & \best{65.2} & \best{54.2} & \best{57.3} & \best{57.1}\gain{3.8} \\
  \midrule
  \rowcolor{tblBand}
  \multicolumn{14}{@{}l}{\grp{MiMo-Audio-7B-Instruct}} \\
  \addlinespace[1.3pt]
  Base & 46.9 & 73.1 & 59.6 & 62.6 & 38.3 & 64.3 & 54.9 & 44.2 & 55.9 & 69.2 & 62.4 & 61.8 & 58.8\phantom{\gain{0.0}} \\
  \rowcolor{tblOurB}
  \ours{SkillFormer} & \best{50.5} & \best{75.2} & \best{62.5} & \best{64.8} & \best{42.0} & \best{66.7} & \best{57.6} & \best{48.6} & \best{58.0} & \best{70.8} & \best{63.7} & \best{63.4} & \best{61.2}\gain{2.4} \\
  \bottomrule
  \end{tabular*}
  \arrayrulecolor{black}
  \vspace{-5pt}
\end{table*}

\vspace{-5pt}
\section{Experiments}
\label{sec:experiments}
\vspace{-5pt}

\vspace{-3pt}
\subsection{Setup}
\vspace{-3pt}

We evaluate SkillFormer on Qwen2.5-Omni-7B~\cite{qwen25},
Kimi-Audio-7B-Instruct~\cite{kimi}, and
MiMo-Audio-7B-Instruct~\cite{mimo}. These models span encoder families
from Whisper-based~\cite{whisper} to self-supervised~\cite{hubert} and
conformer~\cite{conformer}, paired with language backbones derived from
Qwen2~\cite{qwen2}, LLaMA~\cite{llama2}, and
DeepSeek~\cite{deepseekv3}. Other recent models such as Audio
Flamingo~\cite{af2,af3} and MiniCPM-o~\cite{minicpm} could also benefit.
Each adapter uses rank $r{=}16$ and the router has $K{=}6$
adapters with top-2 routing ($n{=}2$). This adds 3.2\% to 3.8\% of the
base model's parameters. Training runs on 4 NVIDIA A100-80G GPUs with
per-GPU batch size~2 and gradient accumulation over 4 steps.

We compare: (1)~\textbf{Base}---the released checkpoint;
(2)~\textbf{SFT}---a single rank-16 LoRA adapter trained on all data;
(3)~\textbf{Multi-LoRA}---all $K{=}6$ adapters active with uniform
$\alpha_k{=}1/K$ and no router; (4)~\textbf{SkillFormer}---our full
method with question-conditioned routing. For MiMo-Audio we report Base
and SkillFormer only.

We evaluate on MMSU~\cite{mmsu}, MMAU-Pro~\cite{mmaupro}, and
MMAR~\cite{mmar}. For MMAU-Pro we use the same closed nonspatial subset
as prior work~\cite{wang2026evoaudio,omnir1}; invalid predictions count as
incorrect.

\subsection{Main Results}

Table~\ref{tab:main} shows consistent improvements across all three
backbones. SkillFormer achieves the highest average on every model,
raising it by 4.1 points on Qwen2.5-Omni, 3.8 on Kimi-Audio, and 2.4 on
MiMo-Audio. Unlike methods that concentrate gains on a single axis,
SkillFormer improves all twelve subcategories on Qwen2.5-Omni, with gains
ranging from 1.7 (MMAR semantic) to 8.0 (MMSU perception).

The progression from SFT to Multi-LoRA to SkillFormer isolates the
contribution of each design choice. SFT with a single adapter gains 1.3
points on Qwen2.5-Omni, showing that even one low-rank update helps.
Multi-LoRA adds six adapters but no routing, gaining another 1.1 points.
SkillFormer introduces question-conditioned routing for a further 1.7
points---the largest single step. The lesson is that having specialized
adapters matters, but letting the model decide which ones to use matters
more.

MMAU-Pro shows the pattern most clearly. Its three categories---speech,
sound, and music---draw on distinct perceptual skills. SFT improves all
three modestly, but Multi-LoRA and SkillFormer widen the gap because
they can route speech queries to one adapter and music queries to another.
On Qwen2.5-Omni, SkillFormer's MMAU-Pro All reaches 60.4, up 4.0 from
Base, the largest single-benchmark gain for this backbone. The pattern
holds on Kimi-Audio, where MMAU-Pro All rises by 4.0 points (49.6 to
53.6), and the weakest subcategory (sound, at 37.1) improves by 5.7
points, the largest single-subcategory gain across all models.

Compared to recurrent reasoning approaches that deepen
computation~\cite{wang2026recurtrace,wang2026aural,cai2026thinking,feng2026coreloop},
SkillFormer offers a complementary advantage: it widens the model's
expertise rather than deepening its analysis. Semantic and cultural
subcategories, which benefit less from deeper iteration, gain 1.7 and
3.1 points respectively on Qwen2.5-Omni under SkillFormer---improvements
that stem from having adapters tuned to language comprehension and world
knowledge alongside those tuned to acoustic perception.

\subsection{Routing Analysis}

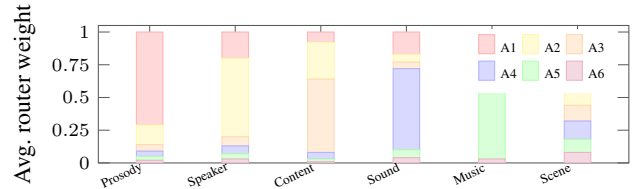
\begin{figure}[t]
\centering
\begin{tikzpicture}
\begin{axis}[
  ybar stacked, bar width=10pt,
  width=\columnwidth, height=3.4cm,
  ylabel style={font=\paperfont\small},
  ylabel={Avg.\ router weight},
  symbolic x coords={Prosody,Speaker,Content,Sound,Music,Scene},
  xtick=data, x tick label style={font=\paperfont\tiny,rotate=25,anchor=east},
  ymin=0, ymax=1.05,
  ytick={0,0.25,0.5,0.75,1.0},
  tick label style={font=\paperfont\scriptsize},
  axis line style={black!65}, tick style={black!65},
  enlarge x limits=0.12,
  legend style={font=\paperfont\tiny,at={(0.98,0.98)},anchor=north east,
    draw=none,fill=white,cells={anchor=west},legend columns=3},
  reverse legend,
]
\addplot[fill=purple!15,draw=purple!30] coordinates
  {(Prosody,0.02)(Speaker,0.03)(Content,0.01)(Sound,0.04)(Music,0.03)(Scene,0.08)};
\addplot[fill=green!15,draw=green!30] coordinates
  {(Prosody,0.03)(Speaker,0.04)(Content,0.02)(Sound,0.06)(Music,0.68)(Scene,0.10)};
\addplot[fill=blue!15,draw=blue!30] coordinates
  {(Prosody,0.04)(Speaker,0.06)(Content,0.05)(Sound,0.62)(Music,0.04)(Scene,0.14)};
\addplot[fill=orange!15,draw=orange!30] coordinates
  {(Prosody,0.05)(Speaker,0.07)(Content,0.56)(Sound,0.05)(Music,0.04)(Scene,0.12)};
\addplot[fill=yellow!20,draw=yellow!40] coordinates
  {(Prosody,0.15)(Speaker,0.60)(Content,0.28)(Sound,0.06)(Music,0.05)(Scene,0.38)};
\addplot[fill=red!15,draw=red!30] coordinates
  {(Prosody,0.71)(Speaker,0.20)(Content,0.08)(Sound,0.17)(Music,0.16)(Scene,0.18)};
\legend{A6,A5,A4,A3,A2,A1}
\end{axis}
\end{tikzpicture}
\vspace{-6pt}
\caption{Router weight distribution by question type on Qwen2.5-Omni-7B.
Each bar shows how the router distributes weight across the six adapters
(A1--A6). Prosody and music questions each concentrate on a single adapter,
while scene questions spread weight more evenly.}
\label{fig:routing}
\vspace{-8pt}
\end{figure}

Fig.~\ref{fig:routing} shows how the router distributes weight across the
six adapters for each question type on Qwen2.5-Omni. Clear specialization
emerges: prosody questions send 71\% of their weight to adapter~1, music
questions route 68\% to adapter~5, and sound event questions concentrate
62\% on adapter~4. Content and speaker queries also lean toward
dedicated adapters but with less extreme sparsity, consistent with these
skills sharing more common ground (both involve speech understanding).
Multi-clip scene queries are the most evenly distributed, drawing on
adapters specialized in different modalities to answer questions that span
them.

This specialization arises without supervision---the skill clusters are
defined by question embeddings, not by hand. The router independently
learns to match questions to the adapter whose cluster best prepared it.
The alignment between clustering and routing validates the approach: skills
that are semantically distinct in question space also benefit from distinct
parameters.

\vspace{-3pt}
\subsection{Ablations}
\vspace{-3pt}

\begin{table}[t]
\caption{Ablations on Qwen2.5-Omni-7B. Each row changes one component of
SkillFormer. Base model Avg.\ is 59.3\%.}
\label{tab:ablation}
\centering
\arrayrulecolor{tblRule}
\paperfont
\setlength{\tabcolsep}{3.5pt}
\renewcommand{\arraystretch}{1.05}
\begin{tabularx}{\columnwidth}{X>{\centering\arraybackslash}p{3.8em}
  >{\centering\arraybackslash}p{4.6em}>{\centering\arraybackslash}p{3.6em}A}
\toprule
\hd{Variant} & \hd{MMSU} & \hd{MMAU-P} & \hd{MMAR} & \hd{Avg.\ (\%)} \\
\midrule
\rowcolor{tblOurB}
\ours{SkillFormer} & 66.5 & 60.4 & 63.3 & 63.4 \\
No router (uniform $\alpha$) & 64.7 & 58.5 & 61.8 & 61.7 \\
Random routing & 63.8 & 57.8 & 61.1 & 60.9 \\
Single adapter (SFT) & 63.2 & 57.5 & 61.0 & 60.6 \\
$K{=}3$ clusters & 65.3 & 59.2 & 62.4 & 62.3 \\
$K{=}12$ clusters & 65.8 & 59.6 & 62.7 & 62.7 \\
No alternating schedule & 64.0 & 58.0 & 61.2 & 61.1 \\
\bottomrule
\end{tabularx}
\arrayrulecolor{black}
\vspace{-8pt}
\end{table}

Table~\ref{tab:ablation} isolates the contribution of each design choice.
Removing the router (uniform $\alpha$) costs 1.7 points, confirming that
question-conditioned routing is the single most valuable ingredient.
Random routing performs even worse ($-2.5$), showing that the router's
decisions carry genuine signal rather than merely adding parameters.

The alternating training schedule matters almost as much: without it,
jointly training all adapters and the router from the start drops accuracy
by 2.3 points. Phase~A gives each adapter time to specialize before
phase~B asks the router to compose them; skipping this step exposes the
adapters to conflicting gradients that flatten their specialization.

Rank sensitivity is mild: changing $r$ from 16 to 8
costs only 0.3 points (Avg.\ 63.1), while $r{=}32$ adds 0.1 (Avg.\ 63.5)
at higher memory cost. Top-2 routing ($n{=}2$) outperforms both
top-1 ($-0.6$) and dense routing ($-0.4$), suggesting that most questions
benefit from blending two expert perspectives but are hurt when too many
adapters dilute the signal.

$K{=}6$ clusters outperform both $K{=}3$ ($-1.1$) and $K{=}12$
($-0.7$). With three clusters, diverse skills are forced to share
parameters. With twelve, each cluster's training set shrinks and the
adapters underfit. Six clusters strike a balance, each large enough to
train a useful adapter and specific enough to avoid interference.

\vspace{-3pt}
\section{Conclusion}
\label{sec:conclusion}
\vspace{-3pt}

We presented SkillFormer, which decomposes audio understanding into
skill-specific low-rank adapters and composes them through a learned
router. The alternating training schedule prevents skill interference, and
the question-conditioned routing ensures each query activates the right
expertise. SkillFormer improves three architecturally distinct LALMs on
three benchmarks, with balanced gains across subcategories. Combining
skill decomposition with richer self-improvement
curricula~\cite{wang2026evoaudio,azr} and adaptive reasoning
depth~\cite{wang2026recurtrace,wang2026aural,cai2026thinking,feng2026coreloop,graves2016adaptive}
is a natural next step.

\balance
\paperfont\sloppy

\bibliographystyle{IEEEbib_etal}
\bibliography{refs}
\end{document}